\documentstyle[preprint,tighten,aps]{revtex}
\begin{document}
\draft
\title{\bf SCHRODINGER EQUATION APPROACH TO NON-LINEAR
$\sigma$-MODELS IN THE LARGE $N$-LIMIT}
\author{ Boris N. Shalaev }
\address{ A.F.Ioffe Physical \& Technical Institute, Russian Academy
of Sciences, \\
194021 St.Petersburg (Russia) \cite{BNS} \\
and Fachbereich Physik, Universitat-Gesamthochschule Essen,\\
D-45117 Essen, Germany }

\maketitle
\begin{abstract}
\noindent

Non-linear $d$-dimensional vector $\sigma$-models are studied
in the large $N$-limit. It is found that a two-point correlation
function obeys a standard Schrodinger equation for a free quantum
particle moving in the $\delta$-function quantum well.
The threshold problem for bound states in this equation
is shown to be equivalent to a critical behavior of these models
above and below the Curie point.

The $SU(N)$- symmetric Ginzburg-Landau (GL) $\sigma$-model subject
to a uniform magnetic field $H$ is considered in the large-$N$ limit
within the Schrodinger equation approach. A upper critical
magnetic field line $H_{c2}(T)$ of type-II superconductors
for an arbitrary external $H$ is obtained without exploiting
the lowest Landau level (LLL) approximation.
Both low-$H$ perturbation expansion terms and exponentially
small corrections to the LLL approximation are calculated.

Correspondences between the one-particle quantum mechanics and
critical phenomena as well as some applications of the above
method to other models of statistical mechanics are also discussed.

\end{abstract}

\vfill
\newpage
\section{ INTRODUCTION }
\renewcommand{\theequation}{1.\arabic{equation}}
\setcounter{equation}{0}

Non-linear vector chiral models such as the $CP^{N}$ and
$O(N), SU(N)$-symmetric ones with a
global non-abelian symmetry are known to possess some remarkable
features, in particular, asymptotic freedom and dynamical mass
generation. These models are an ideal theoretical laboratory for
researchers because they capture all the essentials of
critical phenomena (see \cite{ab1a,ab1b,ab1,ab2,ab3,ab3a} and
references therein).

The standard dynamical approach for treating the models mentioned
in the large-$N$ limit is based on the non-linear Dyson-Schwinger
(DS) equations \cite{ab4}. In this limit the Hartree-Fock
approximation is known to be exact and can be readily solved.
That also holds, for instance, for the $0(N)$-symmetric
$\Phi^{4}$-theory and the Gross-Neveu model
at $N=\infty$ \cite{ab2,ab3,ab4}.

In this paper we will show that in contrast to the DS approach
the two-point correlation function $<S^{*}_{a}(\vec{x})S_{b}(0)>$
of the $O(N), SU(N), CP^{N}$ models in the large $N$-limit
and in the symmetric phase obeys a conventional Schrodinger
equation for a free quantum particle moving in a potential
$(-T)\delta(\vec{x})$ with $T$ being a temperature.

All the well-known results can be obtained in this way. In other
words, statistical mechanics and the elementary quantum mechanics
are shown to be somewhat closely related  subjects, and the relationship
 will be developed here via the Schrodinger equation approach.

It is curious that the Schrodinger equation under discussion is just
the very same toy model discovered long ago for describing dimensional
transmutation phenomena in the non-relativistic quantum mechanics
(see the textbook \cite{ab5}).
As a result, one obtains an amuzing physical interpretation of
dynamical mass generation, critical phenomena, etc. which take place
in non-linear vector $\sigma$-models in terms of quantum mechanics.

Being almost evident from the physical standpoint, nevertheless,
it is somewhat curious to give a precise formulation of
correspondences between quantum field theory and quantum mechanics.

It should be noted that some aspects of this topic have been already
discussed in literature. In the recent paper \cite{ab5a} (see also
references therein) the critical behavior of the binding energy and
of a radius of the bound state $\xi$ near the threshold were considered.
It was also found that the critical exponent $\nu$ of
the correlation length $\xi$ is coincided to that of the
spherical model. The present paper is aimed to develop
the approach presented in \cite{ab5a} and to provide a deeper insight
into correspondences between one-particle quantum mechanics
and critical phenomena.

There is a wide held opinion among experts in the field that a
renormalization-group (RG) approach can be applied only for systems
with an infinite number of degrees of freedom, which lead
 to ultraviolet singularities. Since quantum mechanics
for a free particle doesn't exhibit  ultraviolet
singularities, hence the RG method does not work.

That is wrong because there is a good number of singular
 potentials in quantum mechanics which require a
 short-distance regularization, say, the $\delta$-function
 potential \cite{ab5}. Moreover, as we will see,
just this potential leads to the $\beta$-function coinciding to those
of the non-linear $\sigma$-models in the large-$N$ limit.

Next we study the $SU(N)$-symmetric Ginzburg-Landau (GL) model subject to a
uniform magnetic field. This system is of great interest in both statistical
and condensed matter physics because it is related to the long standing and
challenging problem of the critical behavior of a type-II
superconductor near the upper critical magnetic field $H_{c2}(T)$.
This problem having a rich and long history has been much discussed in
literature (\cite{ab6}).

In brief, it is as follows (\cite{ab7}).
It has been recognized for some time that an external magnetic field
drastically changes the critical properties of superconductors.
The magnetic field hinders the growth of the thermal fluctuations in the plane
perpendicular to ${\bf H}$, since the growth of their correlation length
is restricted by the magnetic length scale $\ell_{H}=\sqrt{hc/eH}$ which is
much shorter than the coherence length $\xi$. This effect of dimensional
reduction results in an enhancement of the longitudinal fluctuations
leading, in particular, to the increase of the lower critical dimension
from 2 to 4.

If critical fluctuations are ignored, the uniform frustration (in the
language of spin models of the vortex lattice) eventually leads to a
continuous phase transition into the Abrikosov flux lattice state.
On the one hand, in contrast to mean-field theory, the standard
renormalization group (RG) approach in $6-\epsilon$ dimensions, in fact,
failed so far to yield insight on the nature of the phase transition due
to the appearence of an infinite number of invariant charges (relevant
scaling variables) inherent to the non-renormalizable scalar $\phi^{4}$
field theory in a field \cite{ab8,ab8a}.

Physical arguments (presented e.g. in \cite{ab6}) support the existence of
a first-order melting transition for the flux lattice.
The conventional $1/N$-expansion, when applied to the
$SU(N)$-symmetric Ginzburg-Landau (GL) model with an $N$-component order
parameter, also gives a first-order phase transition above four dimensions
\cite{ab9}.

On the other hand, however, the results obtained within the framework
of the $1/N$-expansion are somewhat controversial because in more recent
studies a continuous phase transition has been obtained \cite{ab7,ab10,ab11}.

The well-known peculiarity  of the $1/N$-expansion is the symmetry breaking
between particle-hole and particle-particle channels leading
to suppressing the p-p channel and to a unstable
spectrum of fuctuations around the Abrikosov flux solution \cite{ab11a}.

In order to avoid these difficuties and to restore the superconducting
phase transition it was recently
proposed to modify the quartic term in the Ginzburg-Landau Hamiltonian.
Namely, according to \cite{ab11a} the conventional $SU(N)$-symmetric
interaction $(\phi_{a}^{*}\phi_{a})^{2}$ should be replaced by
$2(\phi_{a}^{*}\phi_{a})^{2}- \phi_{a}^{*}\phi_{b}\phi_{a}^{*}\phi_{b}$
being $O(N)\times U(1)$-symmetric.

The question of the nature of the phase transition from
the normal into the mixed state of a type-II superconductor not
discussing here  remains therefore an open problem
(for further discussion, see \cite{ab11a}),
even before the effects of impurities and the
topical question of the vortex glass are to be considered.

It should be also mentioned that in lattice models of
superconductors  the interaction between
thermal fluctuations and the underlying crystal lattice, i.e.
 when a periodic lattice potential is
coupled to the superconducting order parameter,
 can restore the
phase transition into the mixed state  \cite{ab10,ab11,ab12,ab13,ab14}.

The essential ingredient of all the above papers is making use
of the lowest level Landau (LLL) approximation. From a formal point
of view this approach describes a phase diagram of a type-II
superconductor near the upper critical magnetic field $H_{c2}(T)$,
at the best, only for an astronomically large magnetic field as
in a vicinity of the neutron star
${\bf H}\sim H_{0}=\frac{\Phi_{0}}{a^{2}}\sim 10^{5} T$
with $\Phi_{0}=hc/2e$ being the magnetic flux quantum and $a$ being
a lattice spacing. In this unphysical (so-called "ultraquantum" ) limit
one has $\ell_{H} \ll a \ll \xi$.
The problem is whether results obtained within
 the LLL approximation hold for any $H$ or not.

In this paper we wish
to obtain the equation of the phase transition
line $H_{c2}(T)$ in the $SU(N)$-symmetric model
in the large-$N$ limit for an arbitrary magnetic field without
using the LLL approximation. The Schrodinger equation approach
developed below will be exploited.

This paper is organized as follows. In Section II we consider
the  $d$-dimensional $SU(N)$-symmetric non-linear $\sigma$-model
in the large $N$-limit and the conventional Schrodinger
equation for the two-point
spin correlation functions is derived.
Section III develops a RG approach for a free quantum particle
moving in the $\delta$-function quantum well. Correspondences
between statistical mechanics and quantum mechanics are discussed.
Section IV contains a treatment of the GL model subject to
a uniform magnetic field. The discussion of some other models and
of related issues as well as some concluding remarks are contained
in Sec.V.

\section{SCHRODINGER EQUATION FOR THE TWO-POINT CORRELATION FUNCTION}
\renewcommand{\theequation}{2.\arabic{equation}}
\setcounter{equation}{0}

We begin with the $d$-dimensional lattice $SU(N)$-symmetric
spin model,
described by the Hamiltonian

\begin{equation}
H=-J \sum_{<i,j>}[\vec{S}_{i}\vec{S}_{j}^{*}+h.c.]
\label{e1}
\end{equation}
\noindent

with $\vec{S}=(S_{1},...S_{N})$ being a $N$-component complex unit vector
with  the fixed-length constraint

\begin{equation}
\vec{S}(\vec{x})\vec{S}(\vec{x})^{*}=1
\label{e2}
\end{equation}

imposed on spins. Here $<i,j>$ indicates that the summation is over
all nearest-neighboring sites; $J$ being a spin coupling constant.

The standard GL model is a continuous version of the lattice model
Eq.(\ref{e1}) obtained after taking an appropriate continuum limit

\begin{equation}
H=\frac{J}{2}\int d^{d}x|\partial_{\mu}S_{a}|^{2}
\label{e3}
\end{equation}

\noindent
where the summation over $\mu=1,...,d $ is understood.
The partition function associated with eq.(\ref{e3}) reads

\begin{equation}
Z=\int \prod_{a=1}^{N}DS_{a}DS_{a}^{*}\exp(-\frac{H}{T})
\delta(|\vec{S}|^{2}-1)
\label{e4}
\end{equation}

The temperature $T$  has been rescaled so as to include $J$
into the coupling constant $T$. Local fields $S_{a}(\vec{x}),a=1,...N$ are
 not free,because of the constraint Eq.(\ref{e2}).

Now I am concerned with an equation for the two-point correlation
function in the disordered  phase $T>T_{c}$  with the
$SU(N)$ -symmetry being unbroken.

\begin{equation}
G_{ab}(\vec{x})=<S_{a}(\vec{x})S_{b}^{*}(0)>
\label{e5a}
\end{equation}
\noindent

Following the standard textbook \cite{ab3}, one makes use of
the well-known representation for Eq.(\ref{e5a})
in terms of path integral

\begin{eqnarray}
G_{ab}(\vec{x})&=&\frac{1}{Z}\int \prod_{a=1}^{N}DS_{a}DS_{a}^{*}D\lambda
S_{a}(\vec{x})S_{b}^{*}(0)\exp[-A_{1}(\vec{S}(\vec{x}),\lambda(\vec{x}))]\nonumber\\
Z&=& \int \prod_{a=1}^{N}DS_{a}DS_{a}^{*}D\lambda \exp[-A_{1}(\vec{S}(\vec{x}),\lambda(\vec{x}))] \nonumber\\
A_{1}(\vec{S}(\vec{x}),\lambda(\vec{x}))&\equiv&\frac{1}{2T}\int d^{d}x[|\partial_{\mu}S_{a}|^{2}+\lambda(|\vec{S}|^{2}-1)]
\label{e5b}
\end{eqnarray}
\noindent
with $A_{1}(\vec{S}(\vec{x}),\lambda(\vec{x}))$
being an effective action and $\lambda(\vec{x})$ - a Lagrange
multiplier serving to enforce the constraint Eq.(\ref{e2}).

Integrating out fields $\vec{S}(\vec{x})$, one arrives at the equation
\cite{ab3a}

\begin{eqnarray}
G_{ab}(\vec{x})&=&\frac{1}{Z}\int D\lambda
G(\vec{x}, \vec{y}; \lambda)\exp[-A_{2}(\lambda(\vec{x}))]\nonumber\\
Z&=& \int D\lambda \exp[-A_{2}(\lambda(\vec{x}))] \nonumber\\
A_{2}(\lambda(\vec{x}))& = &\frac{1}{2T}\int
d^{d}x[\lambda(\vec{x})-\frac{N}{2} tr \log(-\Delta+\lambda(\vec{x}))]\nonumber\\
\Delta& \equiv &\partial_{\mu}^{2}
\label{e5c}
\end{eqnarray}
\noindent

where

\begin{equation}
G(\vec{x}, \vec{y}; \lambda)=<\vec{y}|\frac{1}{-\Delta+\lambda}|\vec{x}>
\label{e5d}
\end{equation}
\noindent

If a product $TN$ remains finite in the large-$N$ limit
$N \rightarrow \infty$, the path integrals in
Eq.(\ref{e5c}) can be easily evaluated by means of the steepest descend
method \cite{ab3,ab3a}

\begin{equation}
<S_{a}(\vec{x})S_{b}^{*}(0)>= T\delta_{ab} G(\vec{x}, \vec{y}; \lambda_{0})
\label{e5e}
\end{equation}
\noindent

with $\lambda_{0}$ being a saddle point.
The equation for the two-point Green's function under discussion
acquires a simple form \cite{ab1b,ab2,ab3,ab3a}.

\begin{eqnarray}
[-\Delta+m^{2}]G_{ab}(\vec{x})=T\delta_{ab}\delta(\vec{x})
\label{e6}
\end{eqnarray}
\noindent

Here $m^{2} \equiv \lambda_{0}$. The inhomogeneous equation Eq.(\ref{e6})
can be readily transformed into a homogeneous one by making use of
the constraint Eq.(\ref{e2}).
Bearing in mind that $ G_{cc}(0)=1$
the r.h.s of Eq.(\ref{e6}) can be rewritten as
$T\delta_{ab}\delta(\vec{x})=T\delta_{ab}\delta(\vec{x})G_{cc}(\vec{x})$.
As a result, this procedure yields the following equation

\begin{equation}
[-\Delta+m^{2}]G_{ab}(\vec{x})=T\delta_{ab}\delta(\vec{x})G_{cc}(\vec{x})
\label{e7}
\end{equation}
\noindent

Here the Einstein summation convention is assumed.

In the symmetric phase for $T>T_{c}$ it is possible to introduce
an "effective wave function"
$G_{ab}(\vec{x}) \equiv \delta_{ab} \Psi(\vec{x})$.
Substituting this in Eq.(\ref{e7}) we obtain the equation

\begin{equation}
\hat{H}\Psi(\vec{x})=-|E|\Psi(\vec{x})
\label{e9}
\end{equation}
\noindent

where the Hamiltonian $\hat{H}$ is given by

\begin{equation}
\hat{H}=-\Delta-T N \delta(\vec{x});\qquad \qquad |E|=m^{2}
\label{e10}
\end{equation}
\noindent

Eq.(\ref{e9}) is the standard Schrodinger equation for
a free quantum particle moving in the $\delta$-function quantum  well,
moreover the one-particle Green function under discussion
corresponds to the lowest level wave function of
the Hamiltonian Eq.(\ref{e10}).

It is worth noting that Eq.(\ref{e10}) results from a double scaling
limit procedure applied to the model under discussion Eq.(\ref{e1}),
namely, the continuum limit and the large-$N$ limit.

The constraint Eq.(\ref{e2}) giving rise to the potential energy
in Eq.(\ref{e10}) may be regarded as a boundary condition
on $\Psi(\vec{x})$

It is worth noting that the sign of T in the Hamiltonian
Eq.(\ref{e10}) corresponding to the attraction, results from a
compactness of the underlying manifold
( the  sphere $S^{2N-1}=\frac{SU(N)}{SU(N-1)}$ in our case).
In turn, it implies that the quantum well has the bound state.

To avoid confusion let's notice that the Schrodinger
equation Eq.(\ref{e9}) doesn't describe a bound state of two
particles from the fundamental $N$-plet of our model, as it could
be seemed at the first sight, because we consider only the
one-particle Green function Eq.(\ref{e5a}).

\section{ONE-PARTICLE QUANTUM MECHANICS AND CRITICAL PHENOMENA}
\renewcommand{\theequation}{4.\arabic{equation}}
\setcounter{equation}{0}

In this Section we will consider the RG approach in the context
of the Schrodinger  equation with the $\delta$-function potential and will
give precise correspondences between phase transitions in statistical
mechanics and the threshold phenomena in quantum mechanics in spirit
of \cite{ab5}.

At the outset let's find eigenfunctions and eigenvalues of the
Hamiltonian $\hat{H}$ Eq.(\ref{e10}) belonging to the
discrete spectrum.

Suppose for concreteness that $d=2$. The above Hamiltonian $\hat{H}$ being
scale invariant, doesn't contain a mass parameter. Despite this,
the only bound state appears, its wave function and energy value
$|E_{bs}|\equiv m^{2}$ can be
found by applying the Fourier transform to Eq.(\ref{e9})

\begin{equation}
[k^{2}+m^{2}]\psi(\vec{k})=T N \Psi(0)
\label{e11}
\end{equation}
\noindent

where

\begin{eqnarray}
\psi(\vec{k})= \int d^{d}x \Psi(\vec{x})\exp(-i\vec{k}\vec{x})\nonumber\\
\Psi(\vec{x})= \int \frac{d^{d}x}{(2\pi)^{d}} \psi(\vec{k})\exp(i\vec{k}\vec{x})
\label{e11a}
\end{eqnarray}
\noindent

From Eq.(\ref{e11}) it follows that

\begin{equation}
\Psi(0)=TN\Psi(0)\int \frac{d^{2}k}{(2\pi)^{2}(k^{2}+m^{2})}
\label{e12}
\end{equation}
\noindent

This relation links the energy of the bound state $m^{2}$
and the coupling constant $T$ \cite{ab5}.

\begin{equation}
1=TN\int \frac{d^{2}k}{(2\pi)^{2}(k^{2}+m^{2})}
\label{e13}
\end{equation}
\noindent

The integral in the r.h.s. of Eq.(\ref{e13}) is logarithmically
divergent. This reflects the fact that $\delta(\vec{x})$ is the
so-called "singular" potential. Likewise  quantum field theory
it requires introducing a ultraviolet cuttoff $\Lambda=a^{-1}$
with $a$ being a width of the quantum well.

One is led to the equation

\begin{equation}
1=\frac{TN}{2\pi}\log\frac{\Lambda}{m}
\label{e14}
\end{equation}
\noindent

The energy of the bound state is a physical observable quantity
and shouldn't depend on the cuttoff. It implies that the coupling
constant acquires the cuttoff dependence $T(\Lambda)$ in order
to keep $E$ $\Lambda$-independent.

In fact, we can treat Eq.(\ref{e14}) as a simplest example of the
isospectral deformation, i.e. the transform of a potential energy
which doesn't change  the energy spectrum.

It is well known from the theory of solitons and of the inverse
scattering transform
that in the case of 1D Schrodinger equation there is an infinite
dimensional group of isospectral deformations,
being isomorphic to a symmetry group of the
Korteweg-de-Vries equation (\cite{ab15}).

Thus we arrive at the one-loop equation for $T$ being an
isospectral deformation in two dimensions \cite{ab5}

\begin{equation}
\Lambda\frac{dT}{d\Lambda}=-T^{2}
\label{e15}
\end{equation}
\noindent

In the $d$-dimensional space one obtains

\begin{equation}
1=TN(\frac{S_{d}}{(2\pi)^{d}}\frac{\Lambda^{d-2}}{d-2}-K_{d}m^{d-2})
\label{e16a}
\end{equation}
\noindent
where
\begin{eqnarray}
S_{d}&=&\frac{2\pi^{d/2}}{\Gamma(d/2)}\nonumber\\
K_{d}&=&\frac{1}{2^{d}\sin(\pi d/2)\Gamma(d/2)\pi^{(d-2)/2}}
\label{e16b}
\end{eqnarray}
\noindent

where $S_{d}$ is the area of the unit $d$-dimensional sphere.

It is convenient to introduce a dimensionless coupling constant
$t=T\Lambda^{2-d}$. The Gell-Mann-Low equation for $t$ is easily seen to be

\begin{equation}
\Lambda\frac{dt}{d\Lambda}=(d-2)t-t^{2}
\label{e17}
\end{equation}
\noindent

Eq.(\ref{e17}) shows that there exists the critical
value of coupling constant (the fixed point)

\begin{equation}
 T_{c}=\frac{(d-2)(2\pi)^{d}}{N S_{d}}\Lambda^{2-d}
\label{e17a}
\end{equation}
\noindent

Now let's discuss correspondences between the quantum
mechanical approach and the statistical physics description.
On the one hand, we are dealing with the conventional Schrodinger
equation with the singular quantum well. It's solution is well known from
the elementary quantum mechanics.

If a depth of this well is small enough: $T<T_{c}$, the bound
state cannot appear at all. For a deep well $T>T_{c}$ there exists
the only bound state with $E_{bs}=-m^{2}$. In two dimensions
the bound state is known to exist irrespective to a magnitude of
the coupling constant $T>0$.

On the other hand, in the language of statistical mechanics and
quantum field theory, Eq.(\ref{e15}) indicates on asymptotic freedom or
dynamical mass generation at $d=2$ and on a continuous phase
 transition at $d>2$. The ultraviolet stable fixed point of the spherical
model $T_{c}$ corresponds to a threshold value of the coupling
contant in quantum mechanics.

Above $T_{c}$ we have  the $N$-plet of massive scalar bosons,
whilst below $T_{c}$ the Goldstone massless particles appear,
as expected.

Interpretating some features of critical phenomena in terms of
quantum mechanics one may go even further by considering the
critical correlation length $\xi$. It is obvious  that
$\xi^{-1}\sim |E_{bs}|=m$. As can be seen from Eq.(\ref{e16a})
in a close vicinity of the threshold $T_{c}$ the discrete eigenvalue
behaves like $|E_{bs}|\sim |T-T_{c}|^{\frac{1}{d-2}}$.
In the context of statistical mechanics it means that
$\xi\sim \tau^{-\nu}$ where $\nu=\frac{1}{d-2}$ \cite{ab5a}.

The energy gap $E_{bs}(T)$ is a continuous function vanishing
at $T=T_{c}$. It means that the model under discussion undergoes
a second-order phase transition. The first-order phase transition
would correspond to a jump of the gap at a threshold which seems
to be impossible in quantum mechanics.

Above the threshold $T>T_{c}$ the asymptotic behavior of the
 wave function is as follows. The large distance asymptotic
$\xi \ll x$  is given by

\begin{equation}
\Psi(\vec{x})\sim \frac{exp(-\frac{x}{\xi})}{x^{\frac{d-1}{2}}}
\label{e18a}
\end{equation}
\noindent

If the gap vanishes: $m=0$ at $T=T_{c}$, then a wave function of the
ground state exhibits a scale-invariant behavior

 \begin{equation}
 \Psi(\vec{x})\sim x^{2-d}
\label{e18b}
\end{equation}
\noindent

Thus, at the Curie point $T=T_{c}$ the wave function changes
its behavior from exponential at high $T$ to the "power law" at
low $T$. Below the threshold the lowest eigenvalue of $\hat{H}$
 equals zero, hence we arrive again at Eq.(\ref{e18b})
corresponding to the correlation function of massless Goldstone excitations
$<\pi_{a}(\vec{x})\pi_{a}(0)>$ with $\pi_{a}(\vec{x})$
being transverse modes.

All the above correspondences between quantum and statistical \\
mechanics are summarized  at the Table 1.

Before moving to the next topic, we make an observation
concerning critical exponents. It is somewhat surprising that
the $d$-dimensional chiral models in the large $N$-limit taken
at special values of $d$ have some critical exponents
coinciding to those of the 2D Ising model and of the
3-state Potts model on dynamical planar lattices (DPL)
(see Table $II$) \cite{ab21}. To avoid misunderstanding stress
that the models under discussion have quite different values of
critical exponents $\nu$ and $\eta$ (Indeed, two any statistical models,
having different spatial dimensions $d$, cannot have numerically
identical critical exponents, because in that case some hyperscaling
relations containing $d$ would be broken).

\section{SU(N)-SYMMETRIC GINZBURG-LANDAU MODEL IN A UNIFORM MAGNETIC FIELD}
\renewcommand{\theequation}{5.\arabic{equation}}
\setcounter{equation}{0}

Here we shall consider the more interesting and nontrivial
example of the $SU(N)$-symmetric $d$-dimensional
Ginzburg-Landau model Eq.(\ref{e3}) subject to a uniform magnetic
field ${\bf B}$ \cite{ab7,ab16}.

This model is described by the Hamiltonian

\begin{equation}
H=\frac{1}{2}\int d^{d}x|(\partial_{\mu}+i\frac{2\pi}{\Phi_{0}}
A_{\mu})S_{a}|^{2}
\label{e19}
\end{equation}

\noindent
where the vector potential $A_{\mu}$ is
taken within the symmetric gauge

\begin{equation}
{\bf A}=\frac{1}{2} [{\bf B},{\bf r}]
\label{e20}
\end{equation}

\noindent
where ${\bf B}$ is taken along the $z$ axis.
The partition function associated with Eq.(\ref{e19}) reads

\begin{equation}
Z=\int \prod_{a=1}^{N}DS_{a}DS_{a}^{*}\exp(-\frac{H}{T})
\delta(|S|^{2}-1)
\label{e21}
\end{equation}

\noindent
The non gauge-invariant two-point correlation function
of $\vec{S}(\vec{r})$ is given by

\begin{equation}
G_{ab}({\bf r},{\bf r}^{\prime})
=<S_{a}({\bf r})S_{b}^{*}({\bf r}^{\prime})>
\label{e22}
\end{equation}

\noindent
with $G_{ab}({\bf r},{\bf r}^{\prime}) $ being a Green
function of the $d$-dimensional Schrodinger operator

\begin{equation}
[(-i\partial_{\mu}-\frac{2\pi}{\Phi_{0}}A_{\mu})^{2}+m_{0}^{2}]
G_{ab}({\bf r},{\bf r}^{\prime})=T\delta_{ab}\delta({\bf r}
-{\bf r}^{\prime})
\label{e23}
\end{equation}

\noindent
The exact solution of Eq.(\ref{e23}) in an arbitrary gauge reads \cite{ab17}

\begin{eqnarray}
G_{ab}({\bf r},{\bf r}^{\prime})&=&T\delta_{ab}
\exp \left ( -i\frac{2\pi}{\Phi_{0}}
\int_{{\bf r}}^{{\bf r}^{\prime}}dx_{\mu}A_{\mu} \right )
(4\pi)^{\frac{2-d}{2}}\int_{0}^{\infty}du
\frac{\omega u^{\frac{2-d}{2}}}{2\sinh(u\omega/2)} \nonumber \\
&\times& \exp\{-m_{0}^{2}u -\frac{(z-z^{\prime})^{2}}{4u}-\frac{\omega}{8}\coth
(\frac{1}{2}u\omega)[(x-x^{\prime})^{2}+(y-y^{\prime})^{2}]\}
\label{e24}
\end{eqnarray}

\noindent
where $\omega=\frac{2eB}{c}$ is the cyclotron frequency (here we have set
$\hbar=1$ and $2m=1$) and $z$ and $z^{\prime}$ are $(d-2)$-dimensional
longitudinal coordinates. The integral in Eq.(\ref{e24}) is taken over
the straight line connecting the points ${\bf r}$ and ${\bf r}^{\prime}$.

Taking into account the constraint $|\vec{S}(\vec{x})|^{2}=1$ we arrive
at the Schrodinger equation for an effective wave function

\begin{equation}
[(-i\partial_{\mu}-\frac{2\pi}{\Phi_{0}}A_{\mu})^{2}+m_{0}^{2}]
\Psi({\bf r},{\bf r}^{\prime})=TN\delta({\bf r}
-{\bf r}^{\prime})\Psi({\bf r},{\bf r}^{\prime})
\label{e25}
\end{equation}

\noindent

where $\delta_{ab} \Psi({\bf r},{\bf r}^{\prime})\equiv G_{ab}({\bf r},{\bf r}^{\prime})$.

From the field-theoretical point of view we are
dealing with the Abelian Higgs
model defined by the Euclidean Lagrangian, Eq.(\ref{e19}) (or, equivalently,
the $N$-component scalar QED), in a large external magnetic field.

It is important for the problem under consideration that an external
magnetic field effectively reduces the spatial dimension of the system
from $d$ to $d-2$.
Thus, if the spatial dimension of the model Eq.(\ref{e19})
is $4+\epsilon$, then due to the dimensional reduction effect
it actually equals $2+\epsilon$.

At the first glance it looks like the dimensional reduction in
a $O(N)$-symmetric Heisenberg ferromagnet subject to a random
magnetic field. As a matter of fact, the model Eq.(\ref{e19})
differs vastly from a random-field  system because of the lack
both of quenched disorder and of the hidden supersymmetry \cite{ab18}.
Moreover, it will be seen below that the dimensional reduction
is not exact in our case.

In the context of quantum mechanics it is known that an
arbitrary weak uniform magnetic field gives rise to the bound state for
any 3D quantum well \cite{ab19}. In other words, it amounts
to the dimensional reduction.

Now let's exploit the boundary condition for the solution Eq.(\ref{e24})
at the coinciding points $\vec{x}=\vec{x}^{\prime}$:

\begin{equation}
\Psi(\vec{x},\vec{x})=\frac{TN\omega}{(4\pi)^{d/2}}\int_{a^{2}}^{\infty}du\frac{\exp(u\omega/2-m^{2}u)}{2u^{(d-2)/2}\sinh(u\omega/2)}=1
\label{e26}
\end{equation}
\noindent

Notice that in order to obtain Eq.(\ref{e26}) we  have used the
mass renormalization $m_{0}^{2}+\omega/2=m^{2}$.
The integral in the r.h.s. of Eq.(\ref{e26}) is ultraviolet divergent,
a short-distance cuttoff $a^{2}$ to be used. As a result one should
deal with  the cuttoff-dependent coupling constant $T(a)$  so as to
keep the mass independent on $a$:$\frac{dm}{da}=0$ \cite{ab1}.

At criticality the mass term vanishes $m=0$, the equation determining
the upper critical magnetic field $H_{c2}(T)$ reads

\begin{equation}
\frac{1}{T(H)}=\frac{N\omega}{(4\pi)^{d/2}}\int_{a^{2}}^{\infty}du\frac{\exp(u\omega/2)}{2u^{(d-2)/2}\sinh(u\omega/2)}
\label{e27}
\end{equation}
\noindent

Provided $d<4$, the integral at the r.h.s. of Eq.(\ref{e27}) diverges at
large $u$ at the critical point. It implies that in this case
 the $H_{c2}(T)$-line doesn't exist at all.

A straightforward algebra yields the phase boundary
$H_{c2}(T)$ in $d>4$ dimensions:

\begin{equation}
\frac{T(0)}{T(H)}=\frac{d-2}{2}(\frac{H}{H_{0}})^{\frac{d-2}{2}}
\int_{\frac{H}{H_{0}}}^{\infty}dt\frac{t^{(2-d)/2}\exp t }{\sinh t}
\label{e28}
\end{equation}
\noindent

where

\begin{equation}
 T(0)=\frac{2(4\pi)^{d/2}a^{d-2}}{N};\qquad H_{0}=\frac{\Phi_{0}}{a^{2}}
\label{e29}
\end{equation}
\noindent

The properties of the upper critical magnetic field
$H_{c2}(T)$ Eq.(\ref{e28}) are as follows.

First, the phase transition line exists only above 4 dimensions
due to the dimensional reduction.

Second, the characteristic scale of the magnetic field is given by $H_{0}$.
The asymptotic behavior of the r.h.s. of Eq.(\ref{e28}) at the
low-$H$ and at the  high-$H$ regions for $4<d<6$ is given by:

1. at low-$H$ when $ s \equiv \frac{H}{H_{0}}\ll 1$

\begin{equation}
\frac{T(0)}{T(H)}=1+\frac{d-2}{d-4}s+A_{d} s^{\frac{d-2}{2}}- \frac{d-2}{3(d-6)}s^{2}+\frac{d-2}{45(d-10)}s^{4}+0(s^{6})
\label{e30}
\end{equation}
\noindent

where constant $A_{d}$ is given by

\begin{eqnarray}
A_{d}&=&-2\frac{d-3}{(d-4)}
+\frac{d-2}{2}\int_{1}^{\infty}dt t^{\frac{d-2}{2}}\frac{\exp t}{\sinh t}\nonumber\\
&+&\frac{d-2}{2}\int_{0}^{1}dt t^{\frac{d-2}{2}}\frac{t\exp t-\sinh t-t\sinh t}{t\sinh t}
\label{e31}
\end{eqnarray}
\noindent

In fact, Eq.(\ref{e30}) describes a vicinity of the critical
end point $T=T_{c}, H=0$ where several phases coexist and the
 second derivative of $T(H)$ with respect to $H$ diverges
$T^{\prime \prime}(H)\sim H^{(d-6)/2}, H \rightarrow 0$.

2. at high-$H$ for $ 1\ll s$ it is possible to calculate explicitly
 corrections to the LLL-approximation

\begin{equation}
\frac{T(0)}{T(H)}= 2\frac{d-2}{d-4}s+\frac{d-2}{2}\exp (-2s)-\frac{(d-2)^{2}}{8s}\exp(-2s) +0(s^{-2}\exp(-2s))
\label{e32}
\end{equation}
\noindent

It is small wonder that the corrections to the LLL approximation
are exponentially small, because non LLL levels are massive.
In contrast to the first case, provided
$s\rightarrow \infty$ the $H_{c2}(T)$ line goes away to infinity.

Finally, the critical exponent of the correlation
length found from Eq.(\ref{e26}) equals $\nu=\frac{1}{d-4}$ \cite{ab7}.

\section{CONCLUDING REMARKS}
\renewcommand{\theequation}{6.\arabic{equation}}
\setcounter{equation}{0}

Thus, it has been demonstrated that the Schrodinger equation for
a free quantum particle moving in a  $\delta$-function well truly
captures essentials of the critical behavior both above and
below $T_{c}$ for
some non-linear vector $\sigma$-models in the large-$N$ limit
with different global non-abelian symmetries,
namely, $SU(N), 0(N)$ and $CP^{N}$. All these models
in large-$N$ limit are non-trivial systems of interacting Goldstone
particles, undergoing a continuous phase transition at a finite
temperature.
All of them  are described by the beta-function Eq.(\ref{e17})
being to some extent "universal".

Notice that the above approach cannot be extended to both gauge
and matrix models. The reason for that is that in the large-$N$
limit vector non-linear $\sigma$-models are equivalent
to a theory of free massless particles.
This is not the case for gauge and matrix models \cite{ab3a}.

It is interesting that the results obtained can be applied to some
other models of  statistical mechanics.
For instance, within the framework of the $2D$ supersymmetric non-linear
$\sigma$-model suggested by E.Witten in \cite{ab19a} it is easy to show
that in the large-$N$ limit the two-point correlator of the $\vec{n}$
field

\begin{equation}
<\vec{n}(x)\vec{n}(0)>
\label{e36}
\end{equation}
\noindent

also obeys Eq.(\ref{e9}).
It is most remarkable that if $d=2, N=3$,  the one-loop
beta-function of this supersymmetric model is exact!
(\cite{ab19c}).

Another model deserved to be mentioned is
the popular Kardar-Parisi-Zhang model.
Without going into details, mention that considering
the KPZ-equation in the 2-particle sector, one just arrives
at the effective Schrodinger equation Eq.(\ref{e9}) with the
"universal" beta-function Eq.(\ref{e17}) \cite{ab20}.

Concerning the continuous GL model in an external magnetic
field Eq.(\ref{e9}) discussed above, it should be noted that
the above approach doesn't use the standard LLL approximation,
results obtained hold  for an arbitrary magnetic field.

Corrections to this approximations are found to be
exponentially small confirming the conjecture advanced
in \cite{ab8} that the LLL approximation provides a correct
description of critical properties at the close vicinity of
$H_{c2}(T)$.

This model doesn't exhibit a phase transition below 4 dimensions.
In order to suppress the dimensional reduction one might  consider
the same model on a lattice in spirit of the approach developed
in \cite{ab7,ab10,ab12,ab13,ab14}. Within the Schrodinger
equation approach, one gets a remarkable correspondence
between a free Bloch electron moving in the crystall lattice
in a uniform magnetic field with the impurity
$(-T)\delta$-function well and the phase-boundary line $H_{c2}(T)$
for type-II superconductors.

More precisely, it is obvious from the physical standpoint that
that electron can be captured by this well  to create a bound
state around the impurity atom. The threshold for the bound
state level $|E_{bs}|$ is just the upper-critical magnetic field
for a uniformly frustrated model.

In fact, we have got a well
known and understood problem of solid state physics where
 the electron is described by the Azbel-Harper-Hofstadter
Hamiltonian with  the delta-function quantum well
\cite{ab22,ab23,ab24}.
This problem can be solved, at least, numerically in the
commensurate case for rational
values of frustration, i.e. when the magnetic flux through the
plaquette is given by $
\frac{\Phi}{\Phi_{0}}=2\pi \frac{p}{q}$ where $p,q$ are
 mutially prime integers. Such investigations would allow one
to determine the dependence $T_{c}(q); q=1,2,...$.
This model as well as a model with weak quenched disorder
are left for future studies.

\section{Acknowledgements}
This work was supported in part by the Russian Foundation for
Basic Research Grant No.98-02-18299. The author is most
grateful to Fachbereich
Physik Universitat-Gesamthochschule Essen, where this work was done,
for support and exceptionally warm and kind hospitality.
He is much benefitted from numerous interesting and
helpful discussions
with H.W.Diehl, K.J.Wiese, Yu.M.Pis'mak, A.I.Sokolov, S.A.Ktitorov
and A.V.Goltsev. He is indebted to K.J. Wiese
for pointing out the connection between the beta-function
Eq.(\ref{e17}) and the KPZ-model and to M.V.Sadovskii for drawing
his attention to the paper \cite{ab5a}. Finally, I would like to thank
J.Zinn-Justin for his valuable and encouraging remarks.

\newpage
\begin{table}
\begin{tabular}{rccccccc}
\qquad ${\rm QUANTUM \quad MECHANICS}$ & ${\rm STATISTICAL \quad MECHANICS}$ &  \\ \tableline
wave function $\psi(\vec{x})$  &  correlation function $<\vec{S}(\vec{x})\vec{S}(0)^{*}>$ &  \\
quantum well $-T\delta(\vec{x})$  &  coupling constant $T$ &  \\
 bound state   & dynamical mass generation  \\
 energy gap $E_{bs}$  &  reciprocal of the correlation length $\xi^{-1}$ &  \\
 generator of isospectral deformations  &  beta-function $\beta(T)$ &  \\
 threshold $T_{c}$  & fixed (or Curie) point  $T_{c}$ &  \\
 $E_{bs}\sim (T-T_{c})^{\nu}; \nu=\frac{1}{d-2}$  & $ \xi \sim (T-T_{c})^{-\nu}; \nu=\frac{1}{d-2}$; &  \\
 $T>T_{c};\psi(\vec{x})\sim \exp(-\frac{x}{\xi})x^{1-d}$   & $T>T_{c}; <\vec{S}(\vec{x})\vec{S}(0)^{*}>\sim \exp(-\frac{x}{\xi})x^{1-d}$  &  \\
$T \le T_{c};\psi(\vec{x})\sim x^{2-d}$   & $T \le T_{c}$; Goldstone modes $<\vec{S}(\vec{x})\vec{S}(0)^{*}>\sim x^{2-d}$  &  \\
 no analogy  & local order parameter $<\vec{S}(\vec{x})>$ &  \\
 \end{tabular}
\caption{ Correspondences between quantum mechanics and statistical mechanics}
\end{table}

\begin{table}
\begin{tabular}{rccccccc}
${\rm Critical}$ & ${\rm {\sigma}-model }$ & ${\rm {\sigma}-model\ (d=3) }$ & ${\rm {\sigma}- model\ (d=8/3)}$ &  \\
${\rm index}$ & ${\rm in \ d \ dimensions}$ & ${\rm and\ Ising\ model\ on\  a\ DPL }$ & ${\rm
and\ 3-state\ Potts\ on\ a\ DPL}$ &  \\ \tableline

$\alpha$  &$\frac{d-4}{d-2} $ & $-1$ & $-2$ \\
 $\beta$  &$\frac{1}{2} $ &  $\frac{1}{2}$ & $\frac{1}{2}$ &  \\
$\gamma$  &$\frac{2}{d-2} $ &  $2$ & $3$ &  \\
$\delta$  &$\frac{d+2}{d-2} $ &  $5$ & $7$ &  \\
$d\nu$  &$\frac{d}{d-2} $ &  $3$ & $4$ &  \\
  \end{tabular}
\caption{ Critical exponents of the $d$-dimensional $\sigma$-model with
$N=\infty$ ($d=3; 8/3$), and those of the Ising model and of
the $3$-state Potts model on a dynamical planar lattice}
\end{table}


\begin{references}

\bibitem[*] {BNS} Permanent address.
\bibitem{ab1a} Polyakov A.M., {\em Phys.Lett.}
 {\bf B 59}, 79 (1975);
\bibitem{ab1b}  Br{\'e}zin E. and Zinn-Justin J., {\em Phys.Rev.}
 {\bf B 14}, 3110 (1976);
\bibitem{ab1} Kogut J.B., {\em Rev.Mod.Phys.} {\bf 51}, 659 (1979)
\bibitem{ab2}E.Brezin, and S.R.Wadia (Eds),
{\em The large $N$ expansion in quantum field theory and statistical
physics}, World Sientific, Singapore, (1993);
\bibitem{ab3} J.Zinn-Justin,
{\em Quantum field theory and critical phenomena},\\
Third Edition, Clarendon Press, Oxford, (1999);
\bibitem{ab3a} Polyakov A.M., Gauge, fields and strings,
Harwood Academic Publishers, Chur, Switzerland, 1987.
\bibitem{ab4} Wadia S.,{\em Phys.Rev.} {\bf D 24}, 970 (1981)
\bibitem{ab5} Huang K.,{\em Quarks, leptons and gauge fields},
World Scientific, Singapore, (1981);
 Thorn C., {\em Phys.Rev.} {\bf D 6}, 39 (1972)
\bibitem{ab5a} Apenko S.M., {\em J.Phys.A: Math.Gen.}
{\bf 31}, 1553 (1998)
\bibitem{ab6} For useful reviews, see M.Rasolt and Z.Tesanovich,\\
 {\em Rev.Mod.Phys.} {\bf 64}, 709 (1992);\\
G.Blatter, M.V.Feigel'man, V.B.Geshkenbein,
 A.I.Larkin, and V.M.Vinokur, \\
{\em Rev.Mod.Phys.} {\bf 66}, 1125
(1994)
\bibitem{ab7} Jug G., and B.N.Shalaev,
{\em Phys.Rev.} {\bf B 58}, 12404 (1998);
\bibitem{ab8} Br{\'e}zin E., Nelson D.R., and Thiaville A.
{\em Phys.Rev.} {\bf B 31}, 7124 (1985);
\bibitem{ab8a} Newman T., and Moore M.A.,
 {\em Phys.Rev.} {\bf B 54}, 6661 (1996)
\bibitem{ab9} Affleck I., and Br{\'e}zin E., {\em Nucl.Phys.}
 {\bf B 257}, [FS14], 451 (1985)
\bibitem{ab10} S.A.Ktitorov, B.N.Shalaev, and L.Jastrabik,
 {\em Phys.Rev.}
 {\bf B 49}, 15248 (1994)
\bibitem{ab11} Radzihovsky L., {\em Phys.Rev.Lett.}
 {\bf 74 }, 4722 (1995); see also the discussion by
 Herbut I.F., and Tesanovich Z., {\em Phys.Rev.Lett.}
 {\bf 76 }, 4550 (1996),
 and  Radzihovsky L., {\em Phys.Rev.Lett.}
 {\bf 76 }, 4551 (1996)
\bibitem{ab11a} Lopatin A., and Kotliar G., {\em Phys.Rev.}
{\bf B 59}, 3879 (1999)
\bibitem{ab12} Toner J., {\em Phys.Rev.Lett.} {66}, 2523 (1991)
\bibitem{ab13} S.A.Ktitorov, Yu.V.Petrov, B.N.Shalaev, and
 V.S.Sherstinov, {\em Int.J.Mod.Phys.} {\bf B 6},1209 (1992)
\bibitem{ab14} Choi M.Y., and Doniach S., {\em Phys.Rev.}
 {\bf B 31}, 4516 (1985)

\bibitem{ab15} see for instance, Zakharov V.E., Manakov S.V.,
Novikov S.P., and
Pitaevskii L.P., {\em Theory of Solitons.
Inverse Scattering Method},
Nauka, Moscow (1980) (in Russian);

Calogero F., and A.Degasperis,{\em Spectral Transform and Solitons:
Tools to Solve and Investigate Evolution equations}, North Holland,
Amsterdam, (1982);

\bibitem{ab16} Lawrie I.D., and Athorne C., {\em J.Phys: Math.Gen.}
 {\bf A 16}, L587 (1983);

\bibitem{ab17} Feynman R., and Hibbs A., {\em Quantum Mechanics and
 Path Integrals} McGraw Hill Book Company, New York (1965);

\bibitem{ab18} Parisi, and Sourlas N., {\em Phys.Rev.Lett.}
 {\bf 43}, 744 (1979)

\bibitem{ab19} Demkov Yu.N., and Drukarev G.F., {\em ZheTP}
{\bf 49}, 257 (1965)

\bibitem{ab19a} Witten E., {\em Phys.Rev.} {\bf D16}, 2991, (1977);

\bibitem{ab19b} for a review see,
Vainshtein A.I.,V.I., Zakharov, Novikov V.A., and Shifman M.A.,
{\em Physics of elementary particles and atomic nuclei}
{\bf17}, 472 (1986);

Novikov V.A., Shifman M.A., and Zakharov V.I.,
{\em Physics Reports} {\bf 116}, 103 (1984)

\bibitem{ab19c} Novikov V.A., Shifman M.A., Vainshtein A.I.,
and Zakharov V.I.,
{\em Nucl.Phys.} {\bf B229}, 381 (1983);

Novikov V.A., Shifman M.A., Vainshtein A.I., and Zakharov V.I.,
{\em Fortsch.Phys.} {\bf 32}, 585 (1984)

\bibitem{ab20} Wiese K.J., {\em J.Stat.Phys.} {\bf 93}, 143 (1998)

\bibitem{ab21} Kazakov V.A., and A.A.Migdal,
{\em Nucl. Phys.} {\bf B 311}, 171 (1988)

\bibitem{ab22} Harper P.G., {\em Proc.Phys.Soc.London} {\bf A 68},
 874 (1955)

\bibitem{ab23} Azbel M.Ya., {\em Sov.Phys.JETP} {\bf 19}, 634 (1964)

\bibitem{ab24} Hofstadter D.R., {\em Phys.Rev.} {\bf B 14}, 2239
 (1976)

\end{references}
\end{document}